\documentstyle[preprint,aps,epsfig]{revtex}

\newdimen\nude\newbox\chek
\def\slash#1{\setbox\chek=\hbox{$#1$}\nude=\wd\chek#1{\kern-\nude/}}

\def\hz{\hat z}

\begin{document}

\draft

\title{Radiative Energy Loss of High Energy Partons\\ 
  Traversing an Expanding QCD Plasma}

\preprint{BI-TP 98/04,
%
%
          LPTHE-Orsay 98-20}

\author{R.~Baier}
\address{
Fakult\"{a}t f\"{u}r Physik, Universit\"{a}t Bielefeld,
D-33501 Bielefeld, Germany}

\author{Yu.L.~Dokshitzer\footnote{
Permanent address: Petersburg Nuclear Physics Institute, Gatchina 188350, 
St.~Petersburg, Russia}}
\address{INFN, Sezione di Milano, Milan, Italy}

\author{A.H.~Mueller\footnote{Supported in part by the U.S. Department of 
Energy under Grant DE-FG02-94ER-40819}}
\address{Physics Department, Columbia University, New York, NY 10027, USA}

\author{D.~Schiff\footnote{Laboratoire associ\'e du Centre National de 
la Recherche Scientifique}}
\address{LPTHE, Universit\'e Paris-Sud, B\^atiment 211, F-91405 Orsay, France}

\maketitle

\begin{abstract}
We study analytically the medium-induced energy loss of a high 
energy parton passing through a finite size QCD plasma, which is expanding
longitudinally according to Bjorken's model. We extend the 
BDMPS formalism already applied to static media to the case of a quark 
which hits successive layers of matter of decreasing temperature, and we 
show that the resulting radiative energy loss can be as large as 6 times the 
corresponding one in a static plasma at the reference temperature 
$T = T (L)$, which is reached after the quark propagates a 
distance $L$.

\end{abstract}


\section{Introduction}
Recent work\cite{Baier1,Baier2,Baier3,Baier4,Gyulassy,Wang,Zakharov1,Zakharov2}
on medium-stimulated gluon radiation from fast partons traversing (hot and 
cold) QCD matter starts from the assumption, that the properties of 
the medium and its interactions with the energetic quark or gluon projectile
do not change with time, i.e.  the basic 
parameters $\mu$, which is the typical transverse momentum given to the 
parton by a single scattering in the medium, and the parton's mean free 
path $\lambda$ are kept constant in time. It also means
in particular  that the temperature $T$ remains constant 
 during the time the parton is passing through the QCD plasma. 

In this paper we study analytically the propagation of a  quark, of high
energy  $E$, traversing an \underbar{expanding} hot QCD medium, i.e. we 
investigate jet broadening, induced  gluon radiation and the 
resulting radiative energy loss of the quark. 
Thereby we extend the analysis of \cite{Baier1} to the case of 
time-dependent parameters $\mu$ and $\lambda$. We follow BDMPS 
\cite{Baier2,Baier3}, and we take into account our recent work \cite{Baier4},
in which we also show the equivalence of our approach with B. Zakharov's
\cite{Zakharov1,Zakharov2} formulation of the Landau-Pomeranchuk-Migdal 
effect \cite{Landau} for QCD.

For simplicity we consider a high energy quark entering and passing
through a hot QCD medium. We may imagine the medium to be a quark-gluon
plasma produced in a relativistic
central $A-A$ collision, which occurs at $\tau = 0$. 
 At time $\tau_0$ the quark enters the homogeneous plasma at high 
temperature $T_0$, which expands longitudinally with respect to the 
collision axis.
We may consider $\tau_0$ to be the thermalization time.
For most of our results the limit $\tau_0 \rightarrow 0$ can be taken with
impunity. We shall also state results for the more realistic situation
where the quark is produced at $\tau_0 = 0$ \underbar{in} the (not yet
thermalized) medium.
 The quark, for simplicity, is assumed to propagate in the 
transverse direction with vanishing longitudinal momentum, such that its 
 energy is equal to its transverse momentum. On its way through the plasma 
the quark hits layers of matter which are cooled down due 
to the longitudinal expansion. We also assume that the plasma lives long 
enough so that the quark is able to propagate a given distance $L$ within the 
quark-gluon  phase of matter. 

The properties of the expanding plasma are described by the 
hydrodynamical model proposed by Bjorken \cite{Bjorken}.
The parameters $\mu$ and
$\lambda$ depend on  temperature, and therefore on time.
 The main relation is the scaling law
\begin{equation}   \label{eq:1.1}
T^3 \tau^\alpha = {\rm const} , 
\end{equation}
where $\tau$ is the proper time of the expanding medium; at 
rapidity $y = 0$ it coincides with the distance  (time)
the quark has propagated through the plasma. The power $\alpha$, which we 
approximate by a constant, may take values between $\alpha = 0$ 
 and $\alpha = 1$ for an ideal fluid.

Let us state our main results for an expanding medium with $\alpha < 1$.
 As for the static medium the transverse momentum broadening of 
the jet follows the random walk behaviour, namely the characteristic
width $p^2_{\bot W}$ is proportional to the path length $L$. The 
radiative energy loss per unit distance, $- dE/dz$,
  can be as large as 6 (2) times
the corresponding loss in a static plasma at temperature $T=T (L)$.
The number 6 (2) corresponds to the situation where the quark enters
the expanding plasma from  \underbar{outside} 
(is produced \underbar{inside} the plasma).

This paper is organized as follows. In Section II we treat jet broadening
due to multiple scattering in case of an expanding plasma (with $\alpha < 1$),
and we estimate the characteristic width $p^2_{\bot W}$. Section III deals
with the induced  gluon radiation. In Section IV we derive the energy  
loss of a quark and relate it to $p^2_{\bot W}$.
Following Bjorken \cite{Bjorken} we review the main characteristics
of an expanding plasma in Appendix A. The Green function of the 
Schr\"odinger-like equation with the time dependent ``potential'' is 
studied in Appendix B. Integrals which are necessary in calculating the 
energy loss are presented in Appendix C.

\section{Jet broadening in an expanding medium} \label{jet broadening}

In this section we consider a high energy parton propagating through an 
expanding QCD medium. By multiple scattering a transverse momentum is given 
to the parton. In \cite{Baier3} we have summarized the derivation of the 
resulting transverse momentum broadening for the case of a static 
uniform medium. In the following we generalize this derivation taking into 
account the space-time development of the medium. As described in Appendix A
we assume longitudinal expansion.

Because of the evolution of the medium the  parton  propagating 
in the transverse direction, $z$, is affected by the position-dependent 
density of the plasma $\rho(z)$ and the parton cross 
section $d\sigma / d^2 \vec q_\bot (\vec q_\bot , z)$. 

Based on the probabilistic interpretation\footnote{The main difference from 
the static case is the expression for the absorption of the parton along 
its path between points $z_0$ and $z$: $\exp [ - (z - z_0 )/\lambda ]$ for 
the static and $\exp \left[ - \int^z_{z_0} \, dz^\prime \, \rho (z^\prime ) 
\sigma (z^\prime ) \right]$ for the expanding plasma, respectively.}
the master equation for the probability  $f (q^2_\bot ,z)$ 
for a quark to have transverse momentum $\vec q_\bot$ (orthogonal to its
direction of motion) at position $z$ is 
\begin{eqnarray} \label{eq:2.1}
\frac{\partial f (q^2_\bot , z)}{\partial z} = & - & 
\int f (q^2_\bot , z) \rho (z) \frac{d\sigma}{d^2 \vec q^{~\prime}_\bot} 
(\vec q_\bot - \vec q^{~\prime}_\bot , z) d^2 \vec q^{~\prime}_\bot 
\nonumber \\
  & + & 
\int f (q^{\prime 2}_\bot , z) \rho (z) \frac{d\sigma}{d^2 \vec q^{~\prime}
_\bot}
(\vec q^{~\prime}_\bot - \vec q_\bot , z) d^2 \vec q^{~\prime}_\bot . 
\end{eqnarray}
The first term (loss term) accounts for partons which are 
scattered \underbar{out} of the direction $\vec q_\bot , \vec q_\bot 
\rightarrow \vec q^{~\prime}_\bot$, and the second one (gain term) counts 
those partons which are scattered \underbar{into} the direction $\vec q_\bot$
from all other directions $\vec q^{~\prime}_\bot$, $\vec q^{~\prime}_\bot 
\rightarrow \vec q_\bot$. The result given in \cite{Baier3}
is reproduced with $1/\sigma \, d\sigma / d^2 \vec q_\bot$ and the mean free
path $\lambda = 1 / \rho\sigma$  
\underbar{independent of~}$\underline{z}$,
 where $\sigma = \int
d^2 \vec q_\bot \, d\sigma / d^2 \vec q_\bot$.
With a $z$-dependent mean free path 
\begin{equation}\label{eq:2.2}
\lambda (z) = [ \rho (z) \sigma (z) ]^{-1} , 
\end{equation}
(\ref{eq:2.1}) can be written as  
\begin{equation}\label{eq:2.3}
\lambda (z) \frac{\partial f (q^2_\bot , z)}{\partial z} = - f (q^2_\bot , z) 
+ \int \frac{1}{\sigma} \, \frac{d\sigma}{d^2 \vec q^{~\prime}_\bot} 
(\vec q^{~\prime}_\bot , z) f( (\vec q^{~\prime}_\bot - \vec q_\bot )^2 , z) 
d^2 \vec q^{~\prime}_\bot ,
\end{equation}
which can be diagonalized by defining 
\begin{equation}\label{eq:2.4a}
\tilde f (b^2 , z) = \int d^2 \vec q_\bot e^{- i \vec b \cdot \vec q_\bot}
f (q^2_\bot , z) , 
\end{equation}
and
\begin{equation}\label{eq:2.4b}
\tilde V (b^2 , z) = \int d^2 \vec q_\bot e^{- i \vec b \cdot \vec q_\bot} 
\frac{1}{\sigma} \frac{d\sigma}{d^2 \vec q_\bot} (\vec q_\bot , z) . 
\end{equation}
The resulting equation becomes 
\begin{equation}\label{eq:2.5}
\lambda (z) \frac{\partial \tilde f (b^2 , z)}{\partial z} = - \frac{1}{4}
\vec b^2 \tilde v (b^2 , z) \tilde f (b^2 , z), 
\end{equation}
where
\begin{equation}\label{2.6}
\tilde v (b^2 , z) = \frac{4}{\vec b^2} (1 - \tilde V (b^2 ,z)) , 
\end{equation}
and $\tilde V (0 , z ) = 1$. 
As discussed in \cite{Baier2}
in QCD $\tilde v (b^2 , z)$ has no finite limit for $b^2 \rightarrow 0$, 
nevertheless, (\ref{eq:2.5}) may be solved in a logarithmic 
approximation
\begin{equation} \label{eq:2.7}
\tilde v (b^2 , z) \simeq \mu^2 (z) \tilde v  , 
\end{equation}
independent of $\vec b$. As in \cite{Baier2,Baier3}
we introduce the scale $\mu^2$, with $\mu (z)$ representing a 
\underbar{typical} momentum transfer to the parton in a parton-medium
collision, evaluated at position $z$.
An explicit model for the scattering cross section is given by the 
screened ``potential'' \cite{Gyulassy}
\begin{equation}  \label{eq:2.7a}
V (\vec q^{~2}_\bot ) = \frac{1}{\sigma} \, \frac{d\sigma}{d^2 \vec q_\bot} = 
\frac{\mu^2}{\pi ( \vec q^{~2}_\bot + \mu^2)^2}.
\end{equation} 
For ${\vec b^2} \simeq 0$ we get, using (\ref{eq:2.7})
\begin{equation}\label{eq:2.8}
\frac{\partial \tilde f (b^2 , z)}{\partial z} \simeq - \frac{\vec b^2}{4} 
\hat q (z) \tilde f (b^2 , z) , 
\end{equation}
with the (transport) coefficient \cite{Baier3} defined by
\begin{equation}\label{eq:2.9}
\hat q (z) \equiv \frac{\mu^2 (z)}{\lambda (z)} \tilde v \,\, \simeq  \,\,
\rho (z) \int^{1/b^2}_0 d^2 \vec q_\bot \vec q_\bot^{~2} \frac{d\sigma}
{d^2 \vec q_\bot} . 
\end{equation}
The solution of (\ref{eq:2.8}) is
\begin{equation}\label{eq:2.10}
\tilde f (b^2 , z) = \tilde f_0 (b^2 , z_0 ) \exp \left\{ - \frac{b^2}{4}
\int^z_{z_0} dz^\prime \hat q (z^\prime ) \right\} ,
\end{equation}
from which the characteristic width of the distribution 
$f (q^2_\bot , z)$ is deduced
\begin{equation}\label{eq:2.11}
p^2_{\bot W} (z) = \langle q^2_\bot (z) \rangle \equiv \int^z_{z_0}
dz^\prime \hat q (z^\prime ) . 
\end{equation}
For a hot (massless) medium the $z$ dependence of $\hat q (z)$ may be 
determined from the temperature dependence of the expanding fluid, 
$T = T (z)$. The leading term of the high-temperature expansion for 
$\hat q (z)$ in (\ref{eq:2.9})
is determined by the $T$-dependence of the density $\rho (z)$ of the 
medium, 
\begin{equation}\label{eq:2.12}
\hat q (z) = \hat q (z_0) (T / T_0 )^3 . 
\end{equation}
This implies that the medium  undergoes cooling from $T_0$ to $T$ when the 
parton propagates from $z_0$ to $z$. Using Bjorken's model \cite{Bjorken} 
summarized in Appendix A,
we may write (see eq.(\ref{eq:A.6}))
\begin{equation}\label{eq:2.13}
\hat q (z) = \hat q (z_0) \left( \frac{z_0}{z} \right)^\alpha  .
\end{equation}

Let us consider the interesting case of an interacting and expanding plasma, 
i.e. the case $\alpha < 1$. 
Inserting (\ref{eq:2.13}) into (\ref{eq:2.11})
the integration (with $\alpha = {\rm const}$) gives for 
$z = L$ in the limit $z_0 \rightarrow 0$ the random walk 
behaviour
\begin{equation}   \label{eq:2.16}
p^2_{\bot W} (L) = \frac{\hat q (L) L}{(1 - \alpha )} .
\end{equation}
In general the relationship is
\begin{equation}   \label{eq:2.16a}
p^2_{\bot W} (L) =
 {\hat q (L)} L \, \frac{1- ( \, \frac{z_0}{L})^{ 1-\alpha}}{1 -\alpha} ,
\end{equation}
which shows the delicacy of taking the limits $z_0 \rightarrow 0$,
$\alpha \rightarrow 1 $.

In the high temperature phase of QCD matter, we note that $\alpha = 
1 / (1 + \Delta_1 /3 )$ and, since $\Delta_1 = O (\alpha^2_s )$
(cf. eq.(\ref{eq:A.7}) in Appendix A),
\begin{equation}  \label{eq:2.17}
1 - \alpha \simeq \Delta_1 /3 = O ( \alpha^2_s ) .
\end{equation}


\section{Gluon radiation spectrum in an expanding medium} 
\label{gluon radiation}

Here we generalize the derivation of the soft gluon emission spectrum
\cite{Baier2} to the case of an expanding hot medium. As described in the 
Introduction we assume that the fast quark  is produced by a hard
collision outside the medium. 
Let us first start with the key equations - valid in the static case - of 
sect. 4 in \cite{Baier2}, which are re-examined in \cite{Baier4}.

Because of the Landau-Pomeranchuk-Migdal phenomenon \cite{Landau},
the induced spectrum is determined by an interference, essentially
by the gluon emission amplitude at $t_1$,
 $\vec f (\vec b , t_2 - t_1 )$,  evolved in time
to $t_2 > t_1$, and the complex conjugate Born amplitude
$\vec f^{~*}_0 (\vec b )$ for emission at $t_2$.
We keep all the variables unscaled, as we did in the previous section.
The Born $\vec b$-space amplitude for gluon emission is given by
\begin{equation} \label{eq:3.1}
\vec f_0 (\vec b) = - 4 \pi i ( 1 - \tilde V (b^2)) \frac{\vec b}{b^2}
\simeq - i \pi \mu^2 \tilde v \vec b ,
\end{equation}
where we work in the logarithmic approximation (cf. eq.(\ref{eq:2.7})).
The two-dimensional vector structure of $\vec f_0$ and $\vec f$ takes
into account the two polarizations of the emitted gluon.

The induced gluon radiation spectrum (per unit length), in the limit of soft
gluon energy $\omega$ and in the large $N_c$ limit
(cf. eq.(4.24) in \cite{Baier2}), is given by 
\begin{equation} \label{eq:3.2}
\frac{\omega d I}{d\omega d z} = \frac{\alpha_s N_c}{2\pi^2} \,
\frac{1}{L} \, 2 Re \left\{ \int^L_0 \, \frac{dt_2}{\lambda} \, 
\int^{t_2}_0 \, \frac{dt_1}{\lambda}\,
\int \, \frac{d^2 \vec b}{(2\pi)^2} \, \vec f (\vec b , t_2 - t_1 ) \cdot
\vec f^{~*}_0 (\vec b ) \biggl|^{\kappa = 0}_{\kappa} \right\} .
\end{equation}
Instead of the variable $z$ used in the equations for $p_\bot$-broadening 
it is equivalent to use the time variable $t$.
In the large $N_c$ limit the coupling of the quark
emitting a gluon is given
 by $\frac{\alpha_s C_F}{\pi^2}\simeq \frac{\alpha_s N_c}{2\pi^2}$.

 The factor 
$d t / \lambda = \rho\sigma dt$ counts the number of scatterers in the 
medium. The factor $1/L$ appears in (\ref{eq:3.2}) because the spectrum 
is given  per unit length.
$\lambda$ is the mean free path of the quark and
$\kappa = \frac{\lambda \mu^2}{2\omega}$.
The $\kappa$ limits indicated  in (\ref{eq:3.2}) eliminate the
medium independent factorization  contribution.

These characteristic properties are now taken into account to allow the 
natural generalization to the expanding medium. By properly 
specifying the time dependences we rewrite (\ref{eq:3.2}) as
\begin{eqnarray}   \label{eq:3.3} 
\frac{\omega dI}{d\omega dz} & = & 
\frac{\alpha_s N_c}{2\pi^2} \, \frac{1}{L} \, 2 Re \left\{ \int^{\tau_0 + L}
_{\tau_0} \, \frac{dt_2}{\lambda (t_2)} \, \int^{t_2}_{\tau_0} \,
\frac{d t_1}{\lambda (t_1)}   \right.  \nonumber \\
& \times & \left.  \int \, \frac{d^2 \vec b}{(2\pi )^2} \, \vec f (\vec b ; 
t_2 , t_1 ) \cdot
\vec f^{~*}_0 (\vec b , t_2 ) \biggl|^{\omega = \infty  
( \kappa = 0)}_{\kappa} \right\} ,   
\end{eqnarray}
where we assume that the quark hits the medium at time $\tau_0$ and travels
a path length $L$, as discussed in the Introduction.
The gluon propagation from $t_1 \rightarrow t_2$ is 
controlled by a Green function
\begin{equation} \label{eq:3.4}
\vec f (\vec b; t_2 , t_1 ) = \int d^2 \vec b^\prime \, G (\vec b , 
t_2 ; \vec b^\prime , t_1 ) \vec f (\vec b^\prime ; t_1 , t_1 ) . 
\end{equation}
The initial condition is 
\begin{equation}  \label{eq:3.5}
\vec f ( \vec b ; t_1 , t_1 ) = \vec f_0 (\vec b , t_1 ) ,
\end{equation}
which is given by (\ref{eq:3.1}), where now $\mu = \mu(t_1)$. With 
the definition of the coefficient $\hat q (t)$ given in (\ref{eq:2.9})
the emission spectrum is expressed in a rather symmetric form with respect to 
$t_1$ and $t_2$, namely by 
\begin{eqnarray}   \label{eq:3.6}
\frac{\omega dI}{d\omega dz} & = &
\alpha_s N_c \frac{1}{L} \, Re \left\{ \int^{\tau_0 + L}_{\tau_0} \, 
dt_2 \, \int^{t_2}_{\tau_0} \, dt_1 \, \hat q (t_2 ) \hat q (t_1) \right.
\nonumber \\
&\times & \left.  \int \frac{d^2 \vec b}{2\pi} \, \int \, \frac{d^2 
\vec b^\prime}{2\pi}
\, \vec b \cdot \vec b^\prime ~ G (\vec b , t_2 ; \vec b^\prime , t_1 )
\biggl|^{\omega = \infty}_\omega \right\} . 
\end{eqnarray}
In the logarithmic approximation the amplitude $\vec f (\vec b; t_2 , t_1)$, 
and therefore the Green function, satisfies a Schr\"odinger equation for the 
2-dimensional harmonic oscillator (actually with imaginary 
potential) \cite{Baier2,Zakharov2}.
For fixed $t_1$ the equation reads
\begin{equation}  \label{eq:3.7}
i \frac{\partial}{\partial t_2} \vec f (\vec b ; t_2 , t_1 ) = \left[
+ \frac{1}{2\omega} \vec\nabla^{~2}_b - \frac{1}{2} \omega \omega^2_0 (t_2) 
\vec b^{~2} \right] \vec f ( \vec b ; t_2 , t_1 ) ,
\end{equation}
with $\omega^2_0 (t) \equiv i \hat q (t)/\omega$. For an expanding medium the 
frequency of the oscillator $\omega_0 (t)$ is time-dependent. In the 
Bjorken model \cite{Bjorken}
the temperature of the hot medium scales with  time, as given 
in (\ref{eq:1.1}), which translates to 
\begin{equation}   \label{eq:3.8}
\omega^2_0 (t) = \omega^2_0 (\tau_0 ) \left( \frac{\tau_0}{t} \right)^\alpha .
\end{equation}
The explicit expression for the
Green function is derived in Appendix B 
\cite{Kleinert}, and given by eq.(\ref{eq:B.13}). 
In order to perform the $\vec b$-space integrations in (\ref{eq:3.6})
it is convenient to change variables
\begin{equation}   \label{eq:3.9}
z_{i,f} = 2 i \nu \omega_0 (\tau_0 ) \tau_0 \left( \frac{t_{1,2}}{\tau_0}
\right)^{1/2\nu} ,
\end{equation}
with the index $\nu = 1/(2 - \alpha )$, such that $\frac{1}{2} \leq \nu <
1$.
The $\vec b$-space integral is given by 
\begin{eqnarray}  \label{eq:3.10}
I & \equiv &
\int  \frac{d^2\vec b}{2\pi} \, \int \frac{d^2\vec b^\prime}{2\pi} \,
\vec b \cdot \vec b^\prime \, G (\vec b , t_2; \vec b^\prime , t_1 ) 
\nonumber \\
& = & - \frac{1}{\pi} \left[ \frac{2\nu \tau_0}{\omega (2 i\nu\omega_0 
(\tau_0 ) \tau_0 )^{2\nu}}\right]^2
\frac{(z_i z_f)^{2\nu - 2}}{\left[ I_{\nu-1} (z_i ) K_{\nu-1} (z_f) - 
I_{\nu-1} (z_f) K_{\nu-1} (z_i)\right]^2} ,
\end{eqnarray} 
in terms of modified Bessel functions $I_\nu (z)$ and $K_\nu (z)$
\cite{Abram}.
Inserting $I$ and the time dependence of the coefficient $\hat q (t)$ as 
specified in (\ref{eq:2.13}) into the spectrum (\ref{eq:3.6}) a rather simple
expression is obtained
\begin{eqnarray}   \label{eq:3.11}
\frac{\omega d I}{d\omega dz} & = & 
\frac{\alpha_s N_c}{\pi} \, \frac{1}{L} \, \frac{1}{4\nu^2} \nonumber \\
& \times & Re \left\{ \int^{\tau_0 + L}_{\tau_0} \, \frac{dt_2}{t_2}  \,
\int^{t_2}_{\tau_0} \, \frac{dt_1}{t_1} \, \frac{1}{\left[ I_{\nu - 1}
(z_i) K_{\nu-1} (z_f) - I_{\nu-1} (z_f) K_{\nu-1} (z_i) \right]^2}
\biggl|^{\omega = \infty}_\omega   \right\}   .
\end{eqnarray}
If we set
\begin{equation}   \label{eq:3.12}
x_{i,f} = \tau_0 \left( \frac{t_{1,2}}{\tau_0} \right)^{1/2\nu} , 
\end{equation}
and express the function $K_\nu (z)$ in terms of $I_{\pm \nu} (z)$,
(excluding the case $\nu = 1$), we arrive at
\begin{eqnarray}   \label{eq:3.13}
\frac{\omega dI}{d\omega dz} & = &
\frac{\alpha_s N_c}{\pi} \, \frac{1}{L} \, \left[ \frac{2 \sin \pi (\nu - 1)}
{\pi} \right]^2  \nonumber \\
& \times & Re \left\{ \int^{\hat\tau_0}_{\tau_0} \, \frac{d x_i}{x_i} \,
\int^{x_i}_{\tau_0} \, \frac{dx_f}{x_f} \, \frac{1}{\left[ I_{\nu-1} 
(2 i\nu\omega_0x_i) I_{1-\nu} (2 i\nu\omega_0 x_f) - (x_i \leftrightarrow
x_f)\right]^2} \biggl|^{\omega = \infty}_\omega \right\} , 
\end{eqnarray}
where we put $\omega_0 \equiv \omega_0 (\tau_0)$ and $\hat\tau_0 \equiv
\tau_0 (1 + L/\tau_0 )^{1/2\nu}$ for shorter notation. 

In order to compare with our previous result for
the non-expanding plasma \cite{Baier2} we take  $\nu = 1/2$.
The induced spectrum (\ref{eq:3.13}) then becomes
\begin{eqnarray}  \label{eq:3.14}
\frac{\omega dI}{d\omega dz} & = & 
\frac{\alpha_s N_c}{\pi L} \, Re \int^{L+\tau_0}_{\tau_0} \,
\frac{d x_i}{x_i} \, \int^{x_i}_{\tau_0} \, \frac{dx_f}{x_f} \nonumber \\
& \times & \frac{4}{\pi^2 \left[ I_{1/2} ( i\omega_0 x_f ) I_{-1/2}
(i\omega_0 x_i) - (x_i \leftrightarrow x_f) \right]^2}
\Biggl|^{\omega = \infty}_\omega .
\end{eqnarray}
Using $I_{1/2} (z)= \sqrt{\frac{2}{\pi}} {\rm sinh}z/\sqrt z$
 and $I_{-1/2} (z) = \sqrt{\frac{2}{\pi}}
{\rm cosh}z/\sqrt z$ \cite{Abram} gives
\begin{equation}  \label{eq:3.15}
\frac{\omega dI}{d\omega dz} = \frac{\alpha_s N_c}{\pi L} \, Re \int^L_0 \, 
dx_i \, \int^{x_i}_0 \, dx_f \left[ \frac{i\omega_0}{{\rm sinh} (i \omega_0
(x_f - x_i))} \right]^2 \biggl|^{\omega = \infty}_\omega ,
\end{equation}
where we have put $\tau_0 = 0$. The remaining integrals can be performed 
explicitly, 
\begin{eqnarray}   \label{eq:3.16}
\frac{\omega dI}{d\omega dz} & = &
\frac{\alpha_s N_c}{\pi L} \, Re \, \int^L_0 \, \frac{d x_i}{x_i}
\left[ \frac{i\omega_0 x_i}{{\rm tanh} (i\omega_0 x_i)} - 1 \right] 
\nonumber \\
& = & \frac{\alpha_s N_c}{\pi L} \, Re \left\{ \ln \left( \frac{{\rm sinh}
(i\omega_0 L)}{i\omega_0 L} \right) \right\} \\
& = & \frac{\alpha_s N_c}{\pi L} \, \ln \left| \frac{\sin (\omega_0 L)}
{\omega_0 L} \right| . \nonumber
\end{eqnarray}
This is the radiation spectrum in the $N_c \rightarrow \infty$ limit, 
derived and discussed in \cite{Baier2,Baier4} for a hard quark  entering 
the static QCD medium and radiating a soft gluon. 

On can easily go beyond the large $N_c$ limit and the soft gluon
 approximation in (\ref{eq:3.13}) and (\ref{eq:3.16}).
For a particle in an arbitrary colour representation $R$, 
$\omega_0^2$ should be replaced by 
\begin{equation}   \label{eq:3.17}
 \omega_0^2 \frac{N_c}{2 C_R} (1 -x + \frac{C_R}{N_c} x^2 ) ,
\end{equation}
where $x$ is the gluon energy fraction $x = \omega / E$.
In addition the r.h.s. of eqs.(\ref{eq:3.13}) and (\ref{eq:3.16})
should be multiplied by
\begin{equation}   \label{eq:3.18}
 \frac{2 C_R}{N_c} (1 -x + \frac{ x^2}{2} ) 
\end{equation}
for a spin $\frac{1}{2}$ fermion, and by
\begin{equation}   \label{eq:3.19}
 \frac{2 C_R}{N_c} \,\frac {(1 + x^4 + (1- x)^4 )}{2 (1-x)}
\end{equation}
for a spin 1 particle (e.g. gluon, $C_R =  N_c$) \cite{Baier4}.


\section{ENERGY LOSS IN AN EXPANDING MEDIUM} \label{energy loss}

Next we integrate the radiation spectrum eq.(\ref{eq:3.13}) with respect 
to the gluon energy $\omega$ in order to obtain the energy loss
per unit length
\begin{equation}  \label{eq:4.1}
- \frac{dE}{dz} = \int^E_0 \, d\omega \, \frac{\omega dI}{d\omega dz} ,
\end{equation}
where we extend the limit $E \rightarrow \infty$. 
In analogy with the static case \cite{Baier2}
we introduce  new integration variables $x$ and $\hat z$ 
\begin{equation}  \label{eq:4.2}
2 i \nu \omega_0 (\tau_0 ) x_i =  i (1 + i ) x \equiv \hat x ,
\,\, \hat z = \frac{x_f}{x_i} ,
\end{equation}
leading to 
\begin{equation}   \label{eq:4.3}
\omega = \frac{2 \nu^2 \hat q (\tau_0 )}{x^2} \, x^2_i .
\end{equation}
Taking $\tau_0 = 0$ and performing the $x_i$-integration
\begin{equation}   \label{eq:4.4}
\int\limits^{\tau_0 (L/\tau_0 )^{1/2\nu}}_0  \,
d x_i x_i \hat q (\tau_0) = \frac{1}{2} \tau^2_0 \hat q  (\tau_0)
\left( \frac{L}{\tau_0} \right) \equiv \frac{1}{2} \hat q (L) L^2 ,
\end{equation}
the energy loss can be written as
\begin{eqnarray}   \label{eq:4.5}
- \frac{dE}{dz} & = & 
\frac{2 \alpha_s N_c}{\pi} \left[ \frac{2 \nu \sin \pi (\nu - 1)}{\pi}
\right]^2 \, \hat q (L) L  \nonumber \\
&\times &  Re \, \int^1_0 \, \frac{d\hat z}{\hat z} \, \int^\infty_0 \,
\frac{dx}{x^3} \, \frac{1}{\left[ I_{\nu - 1} (\hat x) I_{1 - \nu}
(\hat x \hat z) - I_{\nu - 1} (\hat x \hat z) I_{1 - \nu} (\hat x) 
\right]^2} \biggl|^{x=0}_x     .
\end{eqnarray}
In order to obtain the subtraction term we 
expand the modified Bessel functions around $x = 0$ \cite{Abram},
$I_\nu (z) \simeq \left( \frac{1}{2} z \right)^\nu / \Gamma (\nu + 1 )$. 
This enables us to write (\ref{eq:4.5}) as
\begin{eqnarray}    \label{eq:4.6}
- \frac{dE}{dz} & = &
\frac{2 \alpha_s N_c}{\pi} \, \hat q (L) L \left[ \frac{2 \Gamma (\nu + 1)
\Gamma (2 - \nu ) \sin \pi (\nu - 1)}{\pi} \right]^2     \nonumber  \\
& \times & \int^1_0 \, \frac{d\hat z}{\hat z \left[ \hat z^{\nu - 1} - 
\hat z^{1 - \nu} \right]^2} \, I (\nu , \hat z ) ,
\end{eqnarray}
where  the function $I (\nu , \hat z)$ is defined in
 eq.(\ref{eq:C.1}) and evaluated   in 
Appendix C. 
We integrate over the $\hat z$ variable (see Appendix C) and obtain the 
analytic expression for the energy loss
\begin{equation}  \label{eq:4.7}
- \frac{dE}{dz} = \frac{\alpha_s N_c}{2} \hat q (L) L \left[ 
\frac{2 \Gamma (\nu + 1) \Gamma ( 2 - \nu) \sin\pi (\nu - 1)}{\pi}
\right]^2 I(\nu) , 
\end{equation}
where the function
\begin{equation}    \label{eq:4.8}
I (\nu) = \frac{1}{4 (1 - \nu)^2 (2 - \nu)} ,
\end{equation}
for $\frac{1}{2} \leq \nu < 1$ is derived in 
eqs.(\ref{eq:C.4})-(\ref{eq:C.10}).
Notice that for $\nu = 1/2$ one recovers the energy loss for a quark 
traversing a static medium of size $L$, as discussed in 
\cite{Baier2,Baier3,Baier4}
\begin{equation}    \label{eq:4.9}
- \frac{dE}{dz} \biggl|_{{\rm static}} = \frac{\alpha_s N_c}{12} \, 
\hat q (L) L .
\end{equation}
Eqs.(\ref{eq:4.7}) and (\ref{eq:4.9}) require that the $\omega$-integration
in (\ref{eq:4.1}) be dominated by small $x$ gluons.
These formulas remain true beyond the large $N_c$ limit.
The colour properties of the traversing particle are contained in the
 (transport) coefficient $\hat q (L)$ given in eq.(\ref{eq:2.9}).

Using (\ref{eq:4.7}) and (\ref{eq:4.9}) one finds 
\begin{eqnarray}     \label{eq:4.10}
- \frac{dE}{dz} & = & 
\frac{6\nu^2}{2 - \nu} \left( - \frac{dE}{dz} \biggl|_{{\rm static}} 
\right)  \nonumber  \\
& =  & \frac{6}{(2 - \alpha ) (3 - 2 \alpha)} \left( - \frac{dE}{dz} 
\biggl|_{{\rm static}} \right) , \,\, \alpha =2 - \frac{1}{\nu} .
\end{eqnarray}
In case the quark is produced \underbar{in} the medium, rather than outside,
\begin{eqnarray}     \label{eq:4.11}
- \frac{dE}{dz} & = &
 {2 \nu} \left( - \frac{dE}{dz} \biggl|_{{\rm static}}
\right)  \nonumber  \\
& =  & \frac{2}{2 - \alpha } \left( - \frac{dE}{dz}
\biggl|_{{\rm static}} \right) 
\end{eqnarray}
replaces (\ref{eq:4.10}), where 
$ - \frac{dE}{dz} \biggl|_{{\rm static}}$
also corresponds to a quark produced in the medium,
and it is 3 times the expression given in (\ref{eq:4.9}) \cite{Baier4}.

We notice that the limit $\nu = 1 \, (\alpha = 1)$
for an expanding ideal relativistic plasma can be taken. In this limit the
maximal loss is achieved. It is bigger by a factor 6  for a quark produced
outside ( 2 for inside ) than in the 
corresponding static case. In this comparison 
 the temperature is taken 
 \underbar{after} the expansion. The 
coefficient $\hat q (L) \, = \, \hat q (T (L))$ has to be 
evaluated at the 
temperature $T (L)$ the quark finally ``feels'' \underbar{after} 
having passed the distance $L$ through the medium which during this 
propagation cools down to $T(L)$.
It is remarkable that the initial temperature $T_0$ of the hot medium does 
not enter the formulae (\ref{eq:4.10}) and (\ref{eq:4.11}),
although $T_0$ is actually diverging in 
the limit $\tau_0 \rightarrow 0$.

As a consequence it is straightforward to generalize the 
relationship between energy loss and $p_\bot$-broadening derived in 
\cite{Baier3} for static nuclear matter to the 
case of an expanding plasma. We derive the relationships
for a quark approaching the medium
\begin{equation}    \label{eq:4.12}
- \frac{dE}{dz} = \frac{\alpha_s N_c}{2} \, \frac{1}{(2-\alpha)
(3-2\alpha)} \, L \frac{\partial}{\partial L} p^2_{\bot W}(L) , 
\end{equation}
and for a quark produced in the medium
\begin{equation}    \label{eq:4.13}
- \frac{dE}{dz} = \frac{\alpha_s N_c}{2} \, \frac{1}{2 - \alpha}
 \, L \frac{\partial}{\partial L} p^2_{\bot W}(L) ,
\end{equation}
relating the energy loss per unit distance
in a hot expanding medium  with the typical transverse 
momentum squared (\ref{eq:2.11})
a jet receives in traversing a length $L$ of a
longitudinally expanding plasma. For $\alpha = 0$ the results of 
\cite{Baier4} are reproduced.


\begin{appendix}
\section{ Properties of an expanding plasma} \label{appendix-a}

Here we recall and briefly summarize the main properties of the space-time
evolution of a hadronic fluid, which is produced by highly relativistic 
nucleus-nucleus collisions. We consider a hydrodynamical model and 
 follow Bjorken \cite{Bjorken} in assuming one-dimensional
 longitudinal expansion. 

In order to obtain the  dependence of the fluid's temperature $T = T (\tau)$
on the proper-time $\tau$ we use the conservation 
law for the entropy density $s$, 
\begin{equation}\label{eq:A.1}
d s / d\tau + s / \tau = 0. 
\end{equation}
We take into account the thermodynamic equation for the pressure
\begin{equation}\label{eq:A.2}
dp / dT = s (T (\tau)), 
\end{equation}
and  express $p$ in terms of a monotonically increasing function of 
temperature, $n (T)$, 
\begin{equation}\label{eq:A.3}
p = \frac{\pi^2}{90} n (T) T^4 . 
\end{equation}
Defining the parameter
\begin{equation}\label{eq:A.4}
\Delta_1 \equiv \frac{T}{n(T)} \frac{dn(T)}{dT},
\end{equation}
which we assume to be temperature independent\footnote{Possible
$T$ dependences of $\Delta_1$ are sketched in ref. \cite{Bjorken}}.
It follows from eqs.(\ref{eq:A.1}-\ref{eq:A.4}) 
\begin{equation}\label{eq:A.5}
\frac{dT(\tau)}{d\tau} = - v^2_s \,\,\frac{T}{\tau} ,
\,\, \, \, v^2_s = (3 +\Delta_1)^{-1/2}, 
\end{equation}
with $v_s$  the sound velocity. 
In the approximation  of $v_s = {\rm const}$  eq.(\ref{eq:A.5})
gives
\begin{equation}\label{eq:A.6}
( T / T_0 )^3 = ( \tau_0 / \tau )^\alpha ,\,\,\,\,\,{\rm with}\, \alpha \equiv
3 v^2_s .
\end{equation}
The parameter $\alpha$ is bounded by $0 \leq \alpha \leq 1$, 
where $\alpha = 0$ means constant temperature and a static medium.
 $\alpha = 1$ is  an ideal relativistic plasma. 

In perturbative thermal QCD \cite{Nieto} the parameter $\Delta_1$ turns 
out to be small, indicating small deviations from  ideal gas behaviour.
For the case of a gluon gas 
\begin{equation}\label{eq:A.7}
\Delta_1 = \frac{165}{8} \left( \frac{ \alpha_s}{\pi}\right)^2 (1 + O 
(\sqrt{\alpha_s} )), 
\end{equation}
in terms of the QCD coupling constant $\alpha_s$, which 
at very high temperatures should be evaluated at the scale $T$.

\section{ Green function} \label{appendix-b}

In order to discuss the solution of eq.(\ref{eq:3.7})
we make a logarithmic approximation and assume that 
$\ln (1 / \vec b^2)$ is slowly varying for small
$\vec b^2$. We then have to solve the 
Schr\"odinger equation for a two-dimensional harmonic oscillator with 
time-dependent frequency. Using  variables familiar from quantum mechanics 
the equation is 
\begin{equation}\label{eq:B.1}
i \frac{\partial}{\partial t} \vec f (\vec b, t) = 
\left[ - \frac{1}{2m} \vec\nabla^2_b + \frac{1}{2} m
  \omega^2_0 (t) \vec b^2 \right] \,\,\vec f ( 
\vec b, t), 
\end{equation}
where the mass of the oscillator is identified with the energy of the emitted
gluon, $m \equiv - \omega$, and the time dependence of the frequency is given 
by the power behaviour 
\begin{equation}\label{eq:B.2}
\omega^2_0 (t) = \omega^2_0 (t_0)(t_0 / t )^\alpha , 
\end{equation}
where the parameter $\alpha$ is discussed in Appendix A, and 
\begin{equation}\label{eq:B.3}
\omega^2_0 (t_0) =  \frac{i \hat q (t_0)}{\omega}  .
\end{equation}
The Green function of the Schr\"odinger equation (\ref{eq:B.1}) 
can be  written in the 
form \cite{Kleinert}
\begin{equation}\label{eq:B.4}
G ( \vec b, t; \vec b^
\prime , t^\prime ) = \frac{m}{2\pi i D (t, t^\prime )} \, \exp \,
\left\{ i S_{{\rm cl}} ( \vec b , t; 
\vec b^\prime , t^\prime ) \right\} , 
\end{equation}
where the fluctuation determinant satisfies the homogeneous differential 
equation
\begin{equation}\label{eq:B.5}
\frac{d^2}{d t^2} D (t, t^\prime ) + \omega^2_0 (t) D (t, t^\prime ) = 0 , 
\end{equation}
with the initial conditions
\begin{equation}\label{eq:B.6}
D ( t^\prime , t^\prime ) = 0, \,\,\, \frac{d}{dt} D (t, t^\prime ) |_
{t=t^\prime} = 1 . 
\end{equation}
The classical action $S_{{\rm cl}}$ in (\ref{eq:B.4}) is determined by the 
classical path $\vec b_{{\rm cl}} (t)$ obeying 
\begin{equation}\label{eq:B.7}
\frac{d^2}{d t^2} \vec b_{{\rm cl}} (t) + 
\omega^2_0 (t) \vec b_{{\rm cl}} (t) = 0 , 
\end{equation}
with
\begin{equation}\label{eq:B.8}
\vec b_{{\rm cl}} (t) = 
\vec b \,\,\,\,\, {\rm and} \,\,\,\,\, 
\vec b_{{\rm cl}} (t^\prime ) = \vec b^\prime .
\end{equation}
It follows that
\begin{equation}\label{eq:B.9}
S_{{\rm cl}} ( \vec b , t; 
\vec b^\prime , t^\prime ) = \frac{m}{2} \left[ 
\vec b_{{\rm cl}} (t) \cdot \frac{d}{dt} 
\vec b_{cl} (t) \right] \biggl|^t_{t^\prime}     .
\end{equation}
The explicit solution of (\ref{eq:B.5}) with (\ref{eq:B.2}) is found in 
terms of modified Bessel functions $I_\nu (z)$ and $K_\nu (z)$ \cite{Abram}
to be 
\begin{equation}\label{eq:B.10}
D (t , t^\prime ) = \frac{2 \nu t_0}{[2 i \nu \omega_0 (t_0 ) t_0 ]^{2\nu}}
(z z^\prime )^\nu \left[ I_\nu (z) K_\nu (z^\prime )  - K_\nu (z) I_\nu
(z^\prime )\right] ,
\end{equation}
where we introduce the variables 
\begin{equation}\label{eq:B.11}
z = z (t) \equiv 2 i \nu \omega_0 (t_0 ) t_0 (t / t_0 )^{1/2\nu} , \,
z^\prime = z (t^\prime ) , 
\end{equation}
with
\begin{equation}\label{eq:B.12}
\nu = 1 / (2-\alpha ) , 
\end{equation}
such that $1/2 \leq \nu \leq 1$.

Using the solution to  (\ref{eq:B.7}) and (\ref{eq:B.8}) in (\ref{eq:B.9})
gives  the Green function (\ref{eq:B.4}) as 
\begin{equation}\label{eq:B.13}
G (\vec b , t; \vec b^\prime , t^\prime ) = \frac{i \omega}{2\pi D (t , 
t^\prime )} \, \exp \,
\left\{ \frac{-i\omega}{2 D (t, t^\prime )} \left[ c_1 
\vec b^2 + c_2 \vec b^{\prime 2}
- 2 \vec b \cdot \vec b^\prime \right] \right\} , 
\end{equation}
with the coefficients
\begin{eqnarray}\label{eq.B.14}
c_1 & = & z (z^\prime / z )^\nu \left[ I_{\nu - 1} (z) K_\nu (z^\prime )  +
 K_{\nu - 1} (z) I_{\nu} (z^\prime) \right], \nonumber \\
c_2 & = & z^\prime (z / z^\prime )^\nu \left[ K_\nu (z) I_{\nu -1} 
(z^\prime ) + I_\nu (z) K_{\nu - 1} (z^\prime ) \right]. 
\end{eqnarray}
The  case  $\underline{\nu = 1/2}$ is 
especially easy to handle, and allows a direct comparison with the results 
already obtained in \cite{Baier2,Baier4}.
The variables  given in (\ref{eq:B.11}) become
 $z (z^\prime ) = i \omega_0 t (t^\prime)$
with $\omega_0 \equiv \sqrt{i \mu^2  \tilde v / \lambda\omega}$.
The functions $I_{1/2} (z)$ and $K_{1/2} (z)$ 
are expressed in terms of hyperbolic functions \cite{Abram}, 
so that the determinant (\ref{eq:B.10}) simplifies to 
$$
D (t , t^\prime ) = \frac{1}{i\omega_0} \, \sinh \,
(z - z^\prime ) = \frac{1}{\omega_0 } \, \sin \, \omega_0 
(t - t^\prime)  .
$$
We note that the Green function (\ref{eq:B.13}) is time-translational
invariant for $\nu = 1/2$. It 
correctly reproduces the result of eq.(5.6) in \cite{Baier2}.

\section{ The integrals $I (\nu , \hz )$ and $I(\nu)$} 
             \label{appendix-c}

Here we evaluate the integral 
\begin{eqnarray}\label{eq:C.1}
I (\nu , \hat z) \equiv Re \, \int^\infty_0 \, \frac{dx}{x^3} \left\{
1 - \frac{ \left[ \hat z^{\nu-1} - \hat z^{1-\nu} \right]^2
\left[ \Gamma (\nu) \Gamma (2-\nu)\right]^{-2}}{\left[ I_{1-\nu} (\hat x) 
I_{\nu-1} (\hat x \hat z) - I_{1-\nu} (\hat x \hat z) I_{\nu-1} (\hat x)
\right]^2}\right\} , 
\end{eqnarray}
with $\hat x \equiv i (1 + i ) x$, by using the integration contour
$C_{1, ... 4}$ in the complex $z = (x + iy)$-plane, which we already 
introduced in \cite{Baier2} (see Appendix D), and which  for 
convenience is reproduced here in Fig. 1.
\begin{figure}[h]
\label{fig16}
\centering
\epsfig{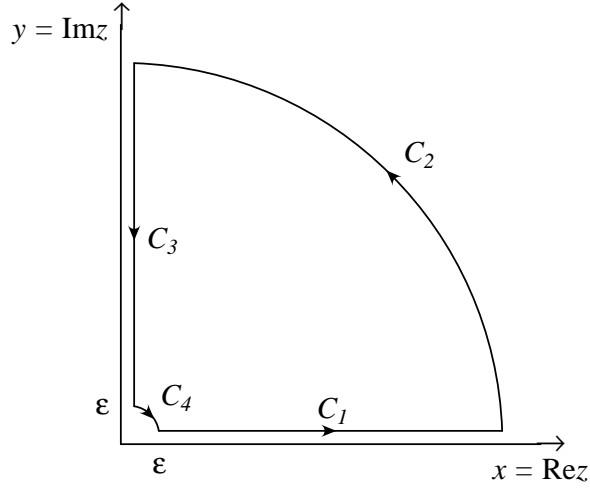}
\caption{\it{Integration contour for the integral $I(\nu , \hat z)$}}
\end{figure}
Following the detailed discussion given in \cite{Baier2}
the integral is performed by calculating the residue of the pole at 
$x = 0$. Since the contribution along $C_2$ vanishes, we find the result 
for $I (\nu , \hat z)$ by adding the contributions from the paths 
$C_1$ and $C_3$, i.e.
\begin{eqnarray}\label{eq:C.2}
I ( \nu , \hat z) & = & 
\frac{1}{2} \left[ I (\nu , \hat z) \right]_{C_1 + C_3} = - \frac{1}{2} 
\left[ I (\nu , \hat z) \right]_{C_4} \nonumber \\
 & = & + \frac{2\pi i}{8} \, {\rm Residue} \, \left[ \frac{1}{x^3} 
\{ \cdot/. \} \right] |_{x = 0} , 
\end{eqnarray}
leading to 
\begin{equation}\label{eq:C.3}
I ( \nu , \hat z) = \frac{\pi\hat z \left[ \left( \frac{1}{2 - \nu}\right)
\left( \hat z^{\nu-2} - \hat z^{2-\nu} \right) + \frac{1}{\nu} 
\left( \hat z^\nu - \hat z^{-\nu} \right) \right] }
{ 4 \left( \hat z^{\nu - 1} - \hat z^{1 - \nu} \right) }.
\end{equation}
For the reader who may not be convinced by the above arguments we
note that we have evaluated the
 integral (\ref{eq:C.1}) numerically using the program
 Mathematica \cite{Mathem}. Stable results were obtained agreeing with
(\ref{eq:C.3}) for a large domain of $z$ and $\nu$,
 $0.1 \leq z \leq 0.8$ and $\frac{1}{2} \leq \nu \leq 0.95$.

In eq.(\ref{eq:4.7}) we stated that the energy loss $- dE / dz$ 
is proportional to the integral $I(\nu)$ defined by
\begin{equation}\label{eq:C.4}
I ( \nu ) \equiv \frac{4}{\pi} \int^1_0 \, \frac{dz}{z} \, 
\frac{I ( \nu , z)}{\left[z^{\nu - 1} - z^{1 - \nu}\right]^2}  =
\int^1_0 \, dz \, \frac{ \frac{1}{(2 - \nu)} \left[ z^{\nu - 2} - z^{2 - \nu} 
\right] + \frac{1}{\nu} \left[ z^\nu - z^{-\nu} \right]}
{\left[ z^{\nu - 1} - z^{1 - \nu} \right]^3}.
\end{equation}
For the special case $\nu = 1/2$
\begin{equation}\label{eq:C.5}
I (1/2) = \frac{2}{3} \, \int^1_0 \, dz = 2/3 .
\end{equation}
For $\nu$ in the interval $\frac{1}{2} \leq \nu < 1$ we evaluate $I(\nu)$ as 
follows. We change the integration variable to 
\begin{equation}\label{eq:C.6}
t =  z^{2(1-\nu)} .
\end{equation}
We regularize the integrand near $t = 1$
by $(1-t)^{-3} \rightarrow (1-t)^{-3+\varepsilon}$, $\varepsilon > 0$,
and arrive at 
\begin{equation}\label{eq:C.7}
I_\varepsilon (\nu) \equiv \frac{1}{2\nu (1 - \nu)(2-\nu)} \, \int^1_0 \,
dt \, 
\frac{\nu \left[ 1 - t^{\frac{1}{1-\nu}+1} \right] + (2 - \nu) \left[
t^{\frac{1}{1-\nu}} - t\right]}
{(1-t)^{3-\varepsilon}},
\end{equation}
where  the limit $\varepsilon \rightarrow 0$
is to be taken  after the integration.
Using the Euler beta-function \cite{Abram} gives
\begin{eqnarray}\label{eq:C.8}
I_\varepsilon (\nu) = & & \frac{1}{2\nu (1-\nu)(2-\nu)} \left\{
\frac{\nu}{\varepsilon -2} - \frac{2-\nu}{(\varepsilon - 1)(\varepsilon -2)}
\right.  \nonumber \\
& - & \left. \frac{\nu (2-\nu) \Gamma (\varepsilon + 1)}{(1-\nu)^2 \varepsilon 
(\varepsilon - 1)(\varepsilon - 2)} \left[ \frac{\Gamma \left( 
\frac{1}{1-\nu} - 
1\right) }{\Gamma \left( \varepsilon + \frac{1}{1-\nu} - 1\right)} - 
\frac{\Gamma \left( \frac{1}{1-\nu}\right)}{\Gamma ( \varepsilon + 
\frac{1}{1 - \nu})}
\right] \right\} . 
\end{eqnarray}
One can easily check that $I_\varepsilon (\nu)$ is regular at $\varepsilon
= 0$. Using \cite{Abram}
\begin{equation}\label{eq:C.9}
\Gamma (z) / \Gamma (\epsilon + z) \mathop{\longrightarrow}_{\varepsilon 
\rightarrow 0} 1 - \varepsilon \psi (z) + O (\varepsilon^2)
\end{equation}
in terms of the digamma function $\psi (z)$, and with the recurrence 
formula $\psi (z+1) = \psi (z) + 1/z$, we finally obtain
\begin{equation}\label{eq:C.10}
I_\varepsilon (\nu) \mathop{\longrightarrow}_{\varepsilon \rightarrow 0}
I (\nu) = \frac{1}{4 (1 - \nu)^2 (2-\nu)} , \,\,\,\,\,\,\,\,
\frac{1}{2} \leq \nu < 1 . 
\end{equation}
This is in agreement with (\ref{eq:C.5}) for $\nu = 1/2$.

\end{appendix}

\vspace{0.5cm} 
\subsection*{Acknowledgements} 
\noindent 
We thank  B.~G. Zakharov for discussions.
R. B. wishes to thank D.~E. Miller for useful comments on the 
interacting gluon gas.
This work is  
supported in part by 
Deutsche Forschungsgemeinschaft (DFG), Contract BA 915/4-2.

\newpage

\end{document}